\documentclass[prl,twocolumn,aps]{revtex4}
\usepackage{graphicx}
\usepackage{amsfonts}

\newcommand{\be}{\begin{equation}}
\newcommand{\ee}{\end{equation}}
\newcommand{\bea}{\begin{eqnarray}}
\newcommand{\eea}{\end{eqnarray}}
\newcommand{\beq}{\begin{equation}}
\newcommand{\eeq}{\end{equation}}
\newcommand{\beqn}{\begin{eqnarray}}

\newcommand{\eeqn}{\end{eqnarray}}

\usepackage{amsfonts}
\usepackage{amssymb}
\usepackage{amsmath}
\usepackage{dsfont}
\usepackage{hyperref}
\usepackage{slashed}
\usepackage{empheq}
\usepackage{multirow}
\usepackage{fancyhdr}
\usepackage{appendix}
\usepackage{subfigure}

\begin{document}

\preprint{MCTP-14-04}

\title{Gravitational Collapse, Chaos in CFT Correlators and the Information Paradox}
\author{Arya Farahi}\email{aryaf@umich.edu}
\affiliation{Michigan Center for Theoretical Physics, Randall Laboratory of Physics, University of Michigan, Ann Arbor, MI 48109, USA}
\author{Leopoldo A. Pando Zayas}\email{lpandoz@umich.edu}
\affiliation{Michigan Center for Theoretical Physics, Randall Laboratory of Physics, University of Michigan, Ann Arbor, MI 48109, USA}

\date{\today}
\begin{abstract}
We consider gravitational collapse of a massless scalar field in asymptotically Anti de Sitter  spacetime. Following the AdS/CFT dictionary we further study correlations in the field theory side by way of the Klein-Gordon equation of a probe scalar field in the collapsing background. We present evidence that in a certain regime the probe scalar field behaves chaotically, thus supporting Hawking's argument in the black hole information paradox proposing that although the information can be retrieved in principle, deterministic chaos impairs, in practice, the process of unitary extraction of information from a black hole. We emphasize that quantum chaos will change this picture.
\end{abstract}

\maketitle

{\it Introduction.} -- The process of black hole formation and evaporation leads to many important conflicts in the interplay between quantum field theory and general relativity. No-hair theorems imply that most information about the collapsing body is lost from the outside region. The discovery that black holes radiate with a perfectly thermal featureless spectrum \cite{Hawking:1974sw} leads to the question of whether the information about the collapsing body is lost with the corresponding loss of unitarity or whether this information is somehow retrievable \cite{Hawking:1976ra}. This conundrum is known as the  black hole information paradox; it  best epitomizes the conflict between quantum field theory and general relativity  and has puzzled researchers for about forty years (see for example \cite{Preskill:1992tc} and more recently its articulation in the language of firewalls \cite{Almheiri:2012rt}).

The information loss paradox is assumed to be implicitly resolved in the context of the AdS/CFT correspondence \cite{Maldacena:1997re} where a gravity theory is equated to an explicitly unitary field theory. In its strictest version, the AdS/CFT correspondence \cite{Maldacena:1997re,Witten:1998qj,Gubser:1998bc,Aharony:1999ti}  states  that string theory in $AdS_5\times S^5$ is equivalent to ${\cal N}=4$ supersymmetric Yang-Mills  with $SU(N)$ gauge group in four dimensions. The implicit resolution of the black hole information paradox in the context of the AdS/CFT \cite{Hawking:2005kf} still leaves us with the daunting question of  {\it how } exactly the paradox gets resolved and by what means information is retrieved from the black hole.

In a recent paper \cite{Hawking:2014tga} Hawking argued that since the gravitational collapse to form an asymptotically AdS black hole will in general be chaotic  the dual CFT on the boundary of AdS will be turbulent, implying, therefore, that information will  be effectively lost, although there would be no loss of unitarity. This situation is standard in deterministic chaos where even though the equations are deterministic there is a practical impossibility to reliably predict the state of the dynamical system after a certain asymptotically large time.

In this manuscript we examine Hawking's claim presented in \cite{Hawking:2014tga} using the standard way the AdS/CFT connects information between the field theory and the dual  gravity. From the main statement of the AdS/CFT correspondence  which is the identification of the field theory and gravity partition functions, it follows that studying the Klein-Gordon (KG) equation with appropriate boundary conditions allows to compute correlations on the field theory side \cite{Witten:1998qj,Gubser:1998bc}. This powerful relation has been improved and generalized  in the conceptual framework of holographic renormalization  \cite{Skenderis:2002wp}. We, therefore, study the KG equation on a gravitationally collapsing background and find various pieces of evidence in favor of chaotic behavior of the scalar field. We show sensitive dependence on the initial conditions practically implying that small uncertainties are amplified exponentially fast leading to the practical impossibility of long-term prediction.

\noindent \emph{Gravitational collapse in asymptotically AdS${}_4$ spacetime}

We consider the dynamics of a massless scalar field $\varphi$  in $4$ dimensions, minimally coupled to gravity with a negative cosmological constant $\Lambda$:
\begin{equation}
\label{Eq:Action}
S=\int d^{4}x\sqrt{-g}\left(\frac{1}{16\pi G}\left(R-\Lambda\right)-\frac{1}{2}(\partial \varphi)^2\right)
\end{equation}
where $G$ is Newton's constant. We focus on spherically symmetric configurations described by the following metric \cite{Bizon:2011gg,deOliveira:2012dt},
{\small
\be
ds^2=\sec^2\left(\frac{x}{\ell}\right)\bigg[-A e^{-2\delta} dt^2 + A^{-1}dx^2 +\ell^2\sin^2\left(\frac{x}{\ell}\right)\,d\Omega^2_{2}\bigg], \label{eq1}
\ee
}
\noindent where $\ell^2=-3/\Lambda$ and  $d\Omega^2$ is the metric on the unit $2$-sphere. The functions $A,\delta$ and the scalar field, $\varphi$, depend on $(t,x)$. The spatial domain is contained in the interval $0 < x <\pi\,\ell/2$. The $AdS$ spacetime, which is the maximally symmetric solution to the vacuum Einstein equations with a negative cosmological constant, $\Lambda$, corresponds to $A=1$, $\delta=0$ and $\varphi=0$.

Introducing the auxiliary variables $\Phi=\varphi^\prime$ and $\Pi=A^{-1}\mathrm{e}^\delta \dot{\varphi}$, where the overdots and primes denote derivatives with respect to $t,x$,  respectively, the field equations read:

\bea
\delta^\prime &=& -4\pi G \ell \cos\left(\frac{x}{\ell}\right)\sin\left(\frac{x}{\ell}\right)(\Pi^2+\Phi^2), \nonumber  \\
A^\prime &=& -4\pi G A\ell \cos\left(\frac{x}{\ell}\right)\sin\left(\frac{x}{\ell}\right)(\Pi^2+\Phi^2),  \nonumber \\
&+ & \frac{1-A}{\ell\cos\left(\frac{x}{\ell}\right)\sin\left(\frac{x}{\ell}\right)}
\bigg[1+2\sin^2\left(\frac{x}{\ell}\right)\bigg] \nonumber  \\
\dot{\Phi}&=&(A\mathrm{e}^{-\delta}\Pi)^\prime, \nonumber \\
\dot{\Pi}&=&\frac{1}{\tan^{2}\left(\frac{x}{\ell}\right)}
\bigg[\tan^{2}\left(\frac{x}{\ell}\right)A\mathrm{e}^{-\delta}\Phi\bigg]^\prime.\label{Eq:Equations}
\eea

\noindent The third equation is a consequence of the definition of the auxiliary variables, and the last is the Klein-Gordon equation $g^{\mu\nu}\nabla_\mu(\partial_\nu \varphi)=0$. Hereafter, we assume units where $4\pi G=1$ and further down we will also fix $\ell=1$.

There is a natural mass function, $m(x,t)$, in AdS${}_4$ spacetime given by

\be
1 - \frac{2 m}{r} + \frac{r^2}{\ell^2} = g^{\alpha\beta}\partial_\alpha r\,\partial_\beta r, \label{eq6}
\ee

\noindent where the standard spherical coordinate $r$ is related to $x$ as  $r = \ell \tan(x/\ell)$. In our case

\be
m(x,t) = (1-A)\frac{\ell \sin\left(\frac{x}{\ell}\right)}{2 \cos^3\left(\frac{x}{\ell}\right)}.\label{eq7}
\ee

\noindent This expression gives the total mass-energy inside a radius $x$ at the instant $t$. The ADM mass of the system is obtained by evaluating the mass function asymptotically, or $M_{\mathrm{ADM}}=\lim_{x \rightarrow \pi\ell/2} m(x,t)$ \cite{deOliveira:2012dt}. The constancy of this quantity is customarily  used in simulations \cite{deOliveira:2012dt,Garfinkle:2011hm} to test the precision of the numerics; we use it here as well.

The fields must satisfy appropriate boundary conditions, in particular \cite{Bizon:2011gg}, near the boundary $x=\pi/2$ , we have ($\rho=\pi/2 -x$):
\bea
\phi(t,x)&=&f_\infty(t)\rho^3 + {\cal O}(\rho^5), \qquad \delta(t,x)=\delta_\infty(t)+{\cal O}(\rho^6), \nonumber \\
A(t,x)&=&1-2M\rho^3 +{\cal O}(\rho^6).
\eea
The AdS/CFT dictionary identifies the asymptotic values of these gravity fields with sources and expectation values for operators in the dual field theory, for example, $M$ above is proportional to the regularized stress energy tensor in the dual field theory  \cite{Skenderis:2002wp}.

To evolve the spacetime we consider a traditional set of initial data \cite{Bizon:2011gg}:
$\Phi(0,x)=0$, $\Pi(0,x) = \epsilon_0 \exp\left(-\tan^2x/\sigma^2\right)$, with $\sigma$ and $\epsilon_0$ as free parameters. Throughout our simulations we will fix $\sigma=0.5$ and consider several values of $\epsilon_0$.

To find the gravitationally collapsing background we solve for, $A, \delta, \Pi$ and $\Phi$ using the boundary conditions explained
\cite{Bizon:2011gg,deOliveira:2012dt}. The initial profile is evolved through time using 4$^{th}$ order Runge-Kutta method until the conditions
for an apparent horizon is satisfied. The numerical integration is stopped at some value of $A_{min}$ but stability of the output is tested
against changing the precise value of $A_{min}$ (see \cite{Garfinkle:2011tc} for a detailed discussion of the methodology). In \cite{Garfinkle:2011hm,Garfinkle:2011tc} this gravitational collapse setup
was used to gain insight into the dual process of thermalization in field theory. In the strictly gravitational context, the
interesting results obtained in \cite{Bizon:2011gg} suggested that AdS spacetime is unstable towards black hole formation in the sense that any arbitrarily small
perturbation leads to the formation of an apparent horizon; a turbulent mechanism for the transfer of energy among the modes was also proposed. In  \cite{deOliveira:2012dt} this turbulent mechanism was quantified by showing that the rate of transfer follows a Kolmogorov-Zakharov spectrum, the mechanism of wave turbulence \cite{ZLKBook} (very different from Kolmogorov 1941) was suggested as the underlying structure.

{\it  Toward CFT correlators  } -- We now consider the massless KG equation for a scalar field in the collapsing
background. Namely, we consider the field $\Psi$  as a probe, that is,
not including its back-reaction on the background; its equation $g^{\mu\nu}\nabla_\mu (\partial_\nu\Psi)=0$, can be written as:

\begin{eqnarray}
\label{Eq:probe}
&-&e^{\delta}\cos^2\left(\frac{x}{\ell}\right)\partial_t \left(e^{\delta}A^{-1}\partial_t \Psi\right) \nonumber \\
&+&e^\delta \frac{\cos(\frac{x}{\ell})}{\sin^2(\frac{x}{\ell})}\partial_x \left(e^{-\delta} A\sin^2(\frac{x}{\ell})\cos(\frac{x}{\ell}) \partial_x \Psi\right) =0.
\end{eqnarray}

It is worth mentioning that the full AdS/CFT dictionary in the context of arbitrary time-dependent configurations has not been rigorously formulated
yet. The natural working assumption, however, is that sources and responses in the field theory are read from the asymptotic behavior
of the field $\Psi$. Spontaneous symmetry breaking is implemented through a boundary condition with no source. A
source is difficult to implement numerically because it corresponds to a non-normalizable mode. For numerical expediency and given
the hyperbolic nature of the
equation versus its more generic elliptic nature in time-independent situations of the original AdS/CFT prescription, we choose
to evolve an initial profile of the probe scalar field of the form: $\Psi(t=0, x)=\sin^3 (x),\,\, \dot{\Psi}(t=0, x)=0 $.  Perhaps due to
the hyperbolic nature of the PDE, we found that using the forward Euler method for integration makes the solution unstable leading to divergencies
fairly rapidly, a better result can be achieved by switching to the backward Euler method which proved to be stable in this case. We settled for fourth order Runge-Kutta which provides the best convergence\footnote{We thank J. Liu for an important discussion on this point.}.

{\it  Sources of chaotic behavior } --
Given that equation \ref{Eq:probe} is linear in the field $\Psi$ we are not going to look for chaotic behavior as sensitivity to the initial conditions
in the evolution of $\Psi$, although linear chaos is certainly a possibility \cite{LinearChaos}. We will study the response of $\Psi$ to a slight
change in the initial conditions that trigger gravitational collapse.
For example, we consider collapse of the Einstein-scalar field system governed by equations \ref{Eq:Equations} with an initial Gaussian profile with amplitudes $\epsilon_0=0.1$ and $\epsilon_0=0.2$. For these two nearby backgrounds we study the probe scalar equation $\Psi$. This protocol will be equivalent to asking, in the field theory side, whether
correlation functions extracted during a thermalization process remain
predictably close so as to permit full reconstruction even after accounting for a slight uncertainty in the amount of initially injected energy.

\begin{figure}[htp]
\begin{center}
\includegraphics[width=1.68in]{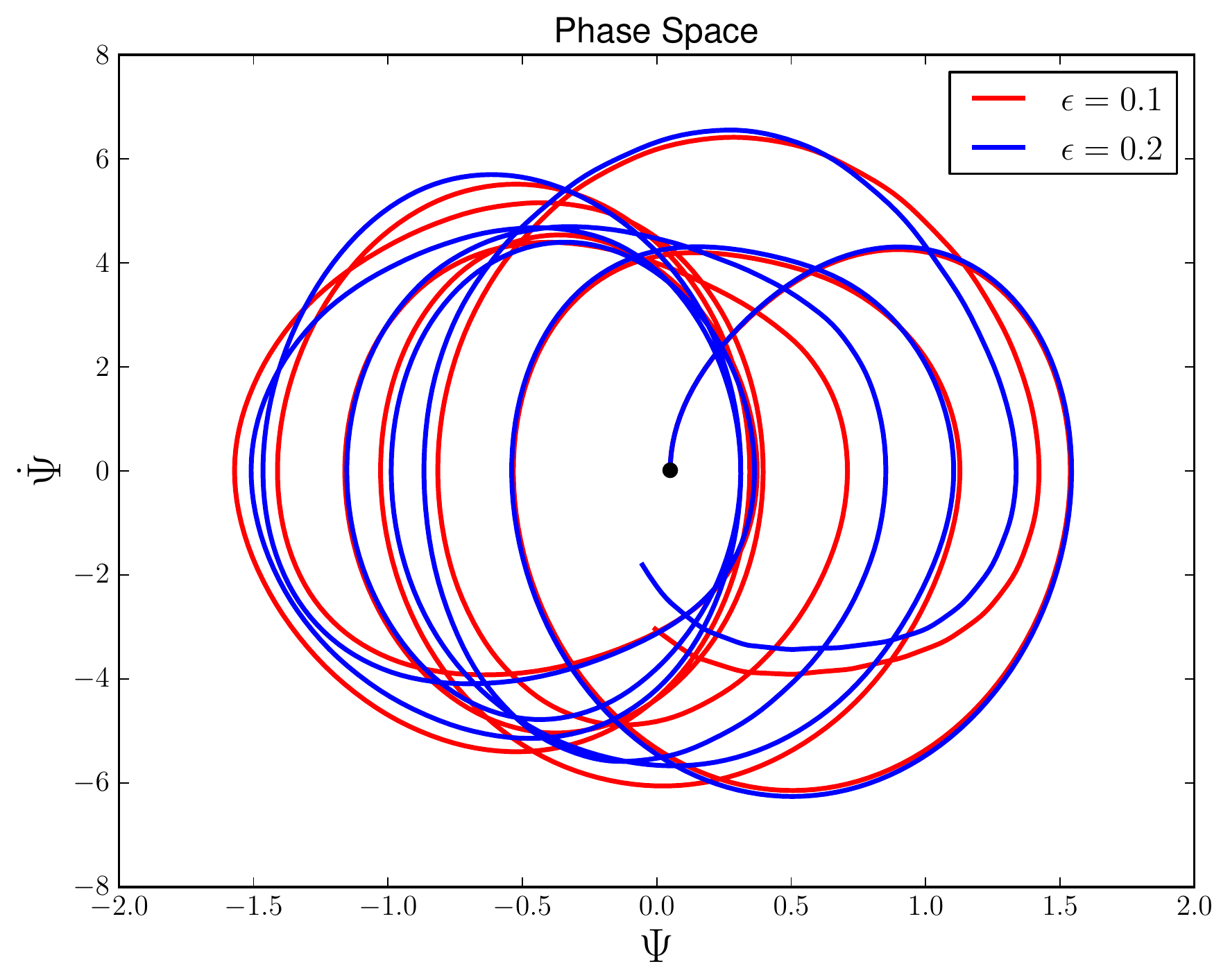}
\includegraphics[width=1.68in]{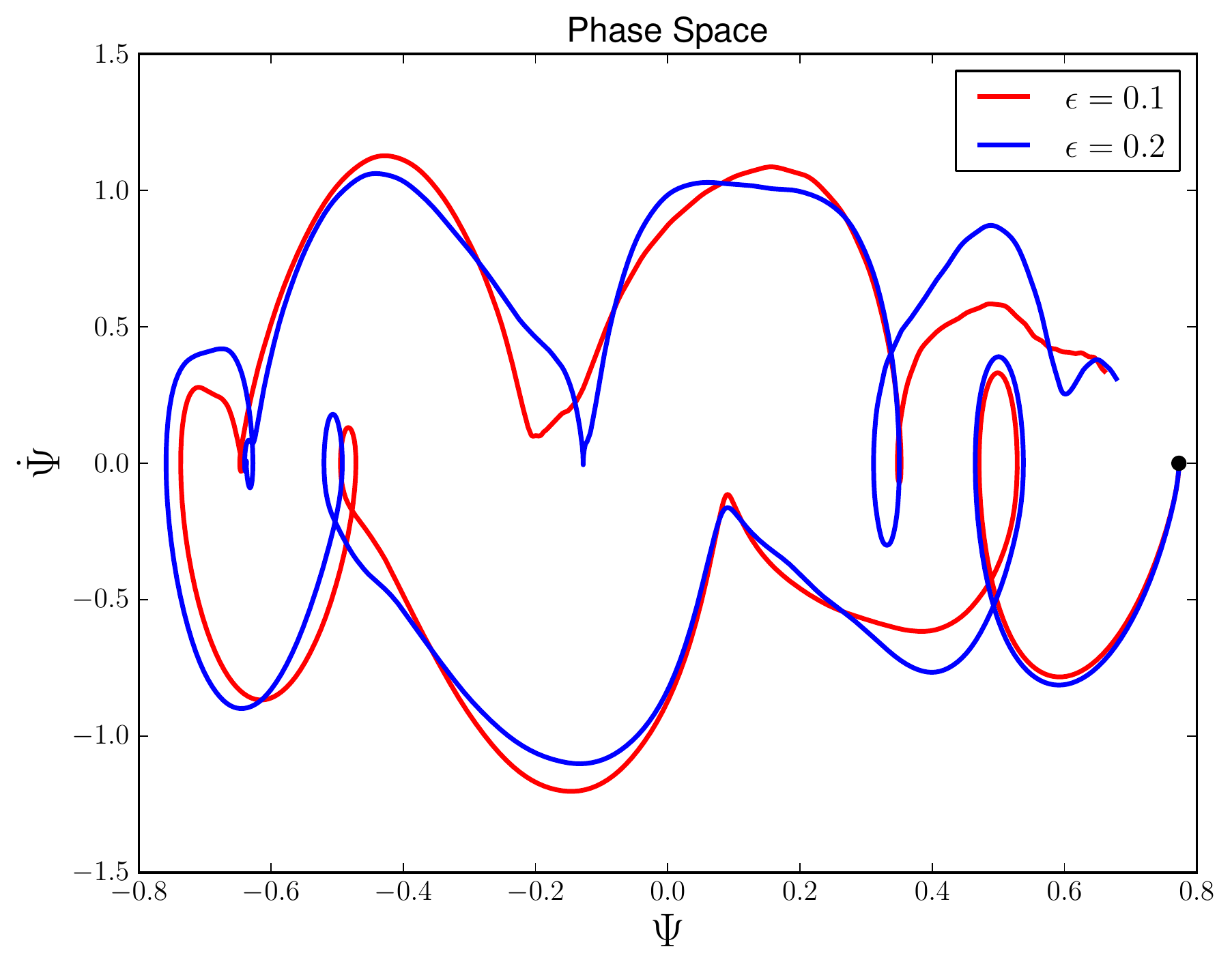}\\
\includegraphics[width=1.68in]{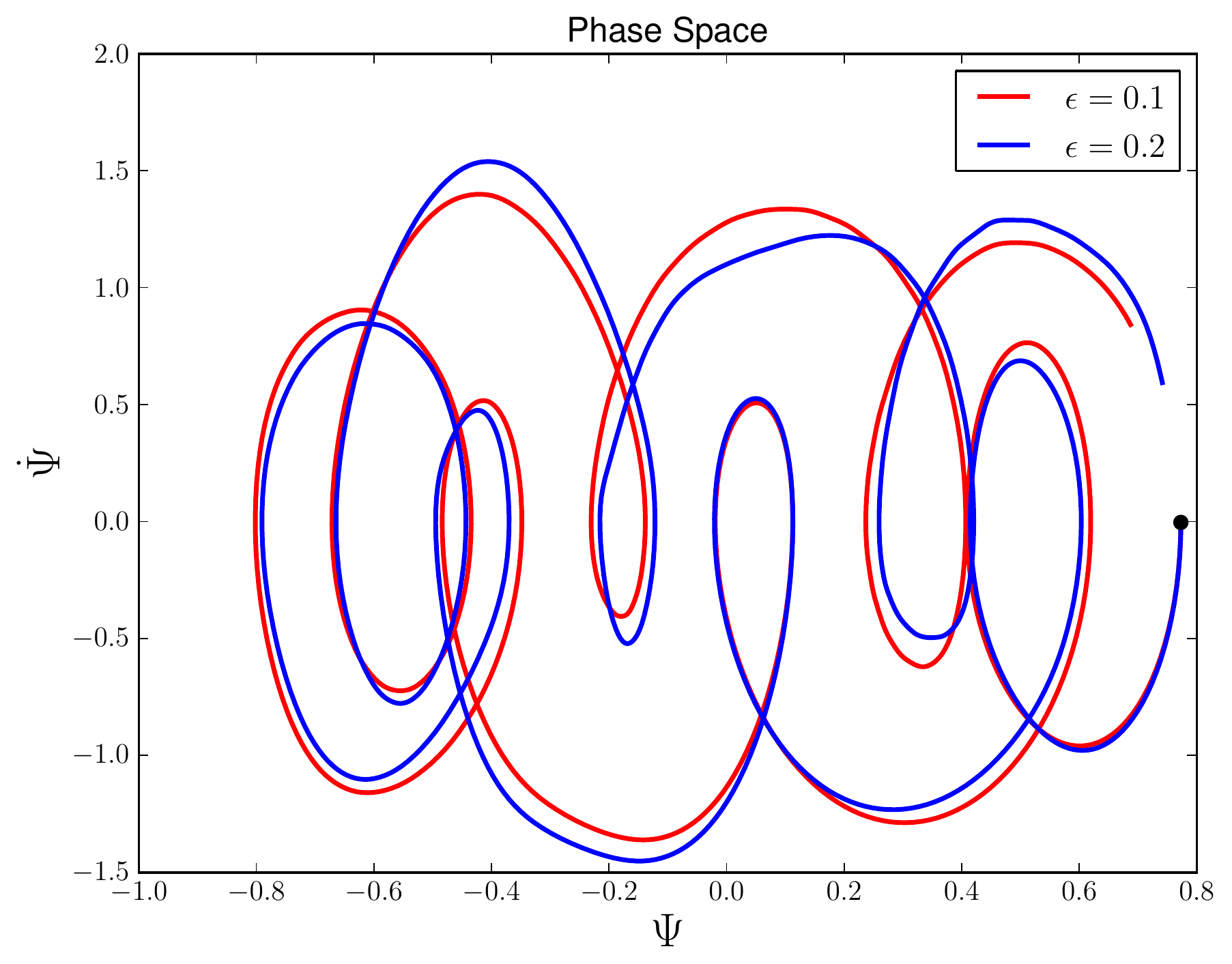}
\includegraphics[width=1.68in]{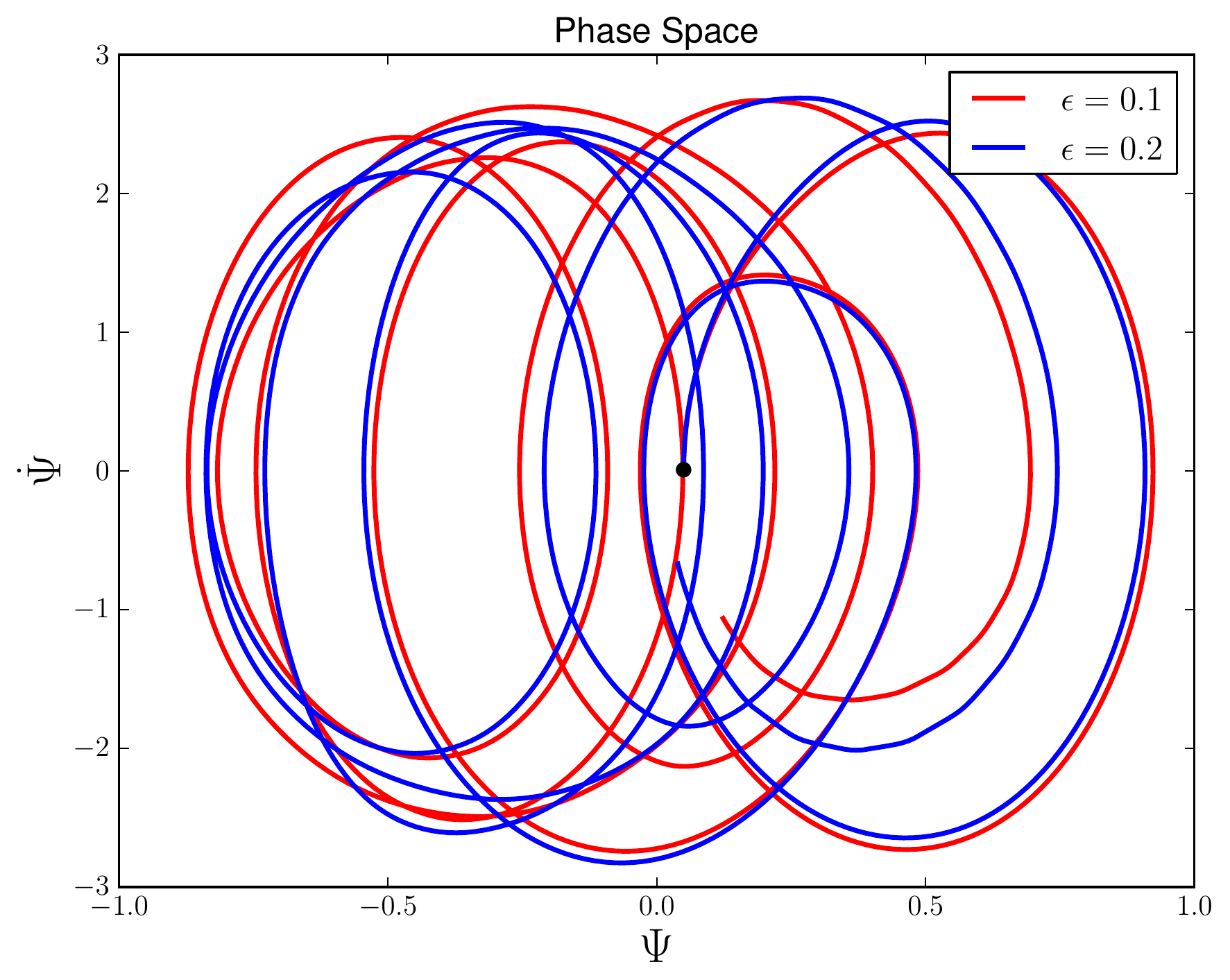}
\caption{\label{PhaseSpace} Phase space values for $\Psi(t,x_i)$ and $\dot{\Psi}(t,x_i)$ corresponding to two simulations of gravitational collapse
with Gaussian profile and amplitudes $\epsilon=0.1, 0.2$. The plots correspond to, from top to bottom and left
to right, $x_i=\pi/16, 3\pi/16, 5\pi/16, 7\pi/16$.}
\end{center}
\end{figure}

It is worth remarking, as eloquently stated in  \cite{Strogatz,Ott}, that no definition of the term chaos is universally accepted. Generically
to prove that a system is chaotic it is  require to show
that the system exhibits sensitive dependence on initial conditions. There are various indicators of
such sensitivity and we consider three of them in what follows.

In figure \ref{PhaseSpace} we follow, in phase space, the evolution of the probe scalar field, $\Psi$. We consider various spatial points
ranging from points behind the eventual apparent horizon for these gravitational collapse simulations to points close to the asymptotic boundary. Chaotic
behavior is clear for all the other points, in particular for $x=7 \pi/16$
which is the closest to the boundary (from where  the field theory
data should be read).  It would be interesting to further study whether
points ``behind'' the  eventual apparent horizon indeed evolve fundamentally differently.  Graphically, figure \ref{PhaseSpace} suggests generically
chaotic evolution of the scalar field $\Psi$. We have performed simulations for various values of the Gaussian profile amplitue. We have also obtained similar results for collapse triggered by a sum of eigenvalues profile which point to certain universality of the results (see discussion in \cite{deOliveira:2012dt}).

\begin{figure}[htp]
\begin{center}
\includegraphics[width=3.3in]{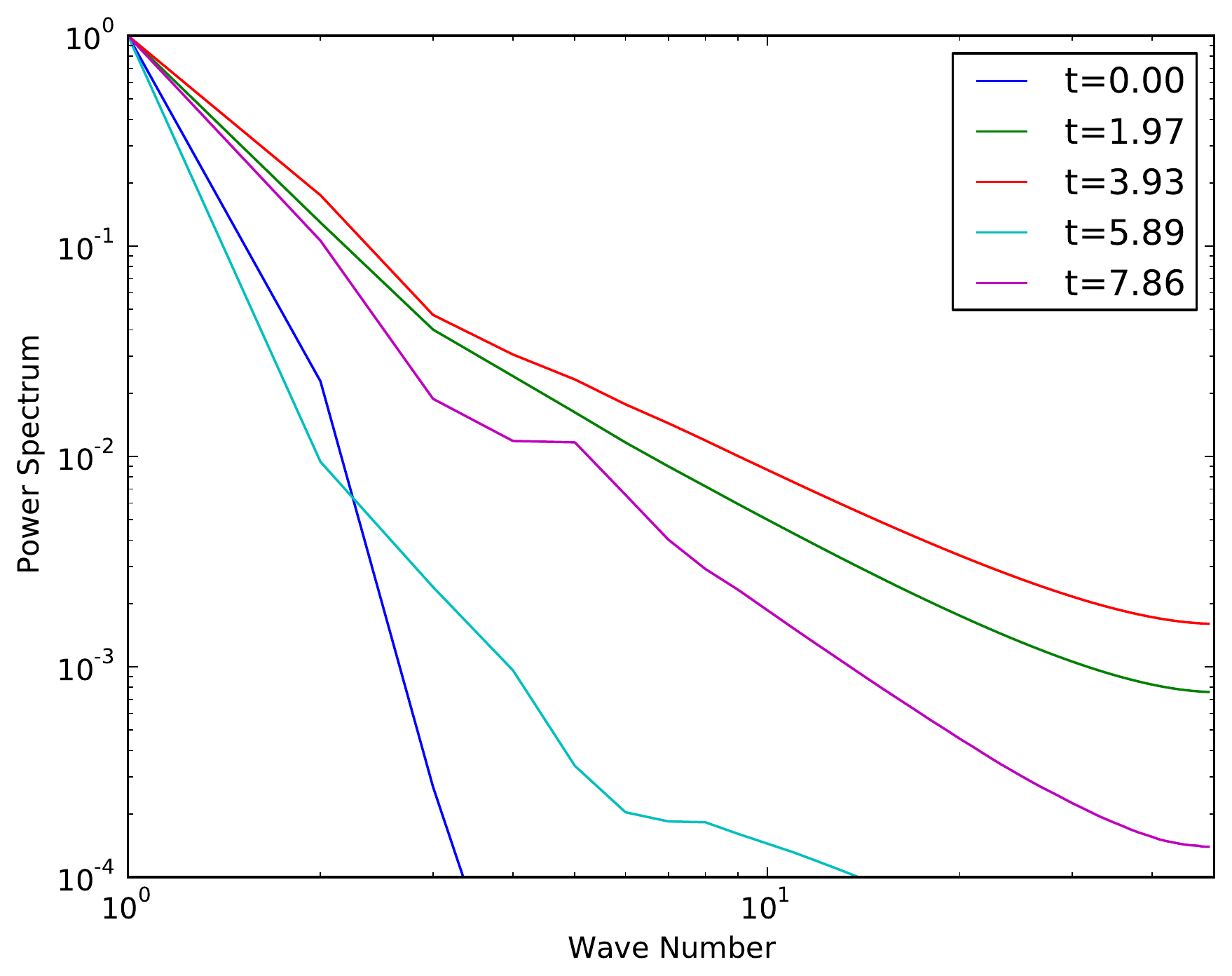}
\caption{\label{PS} For a collapsing simulation with amplitude $\epsilon=0.1$ we plot the power spectrum at various times. We note that an
initially narrow profile widens in the space of frequencies. }
\end{center}
\end{figure}

Next, we turn to a study of the spatial distribution of the initial profile $\Psi(t=0,x)$ as a function
of time; we are interested in the
spatial distribution  after
certain large time $T$, that is, $\Psi(t=T, x)$. At time equal zero we consider a fairly narrow, in spatial frequencies,
profile $\sin^3(2x)$. In figure  \ref{PS} we plot the normalized power spectra for the initial profile
and after evolving the
scalar field $\Psi$ for some large time, $T$. The crucial point is that we start with a narrow profile, note that the power spectrum vanishes $(10^{-5})$ for large frequencies. The late time spectral analysis, close to the formation of an apparent horizon, indicates a profile with many more frequencies activated (order $10^{-2}$ in a wide range). The oscillatory behavior of the power spectrum is inherited from the bounces in the collapsin background. This property of the power spectrum is generically indicative of sensitivity to initial conditions; its trademark example is the H\'enon-Heiles system where starting with two
frequencies a continuum of frequencies is generated.

It is worth pointing out that the power spectrum reported in \cite{deOliveira:2012ac} and later corroborated in \cite{Buchel:2012uh} refers to the
gravitational background and should not be confused with the analysis presented here. In those papers the focus of the analysis was
the time-series of certain quantities such as the scalar Ricci scalar and the local mass.

\begin{figure}[htp]
\begin{center}
\includegraphics[width=3.3in]{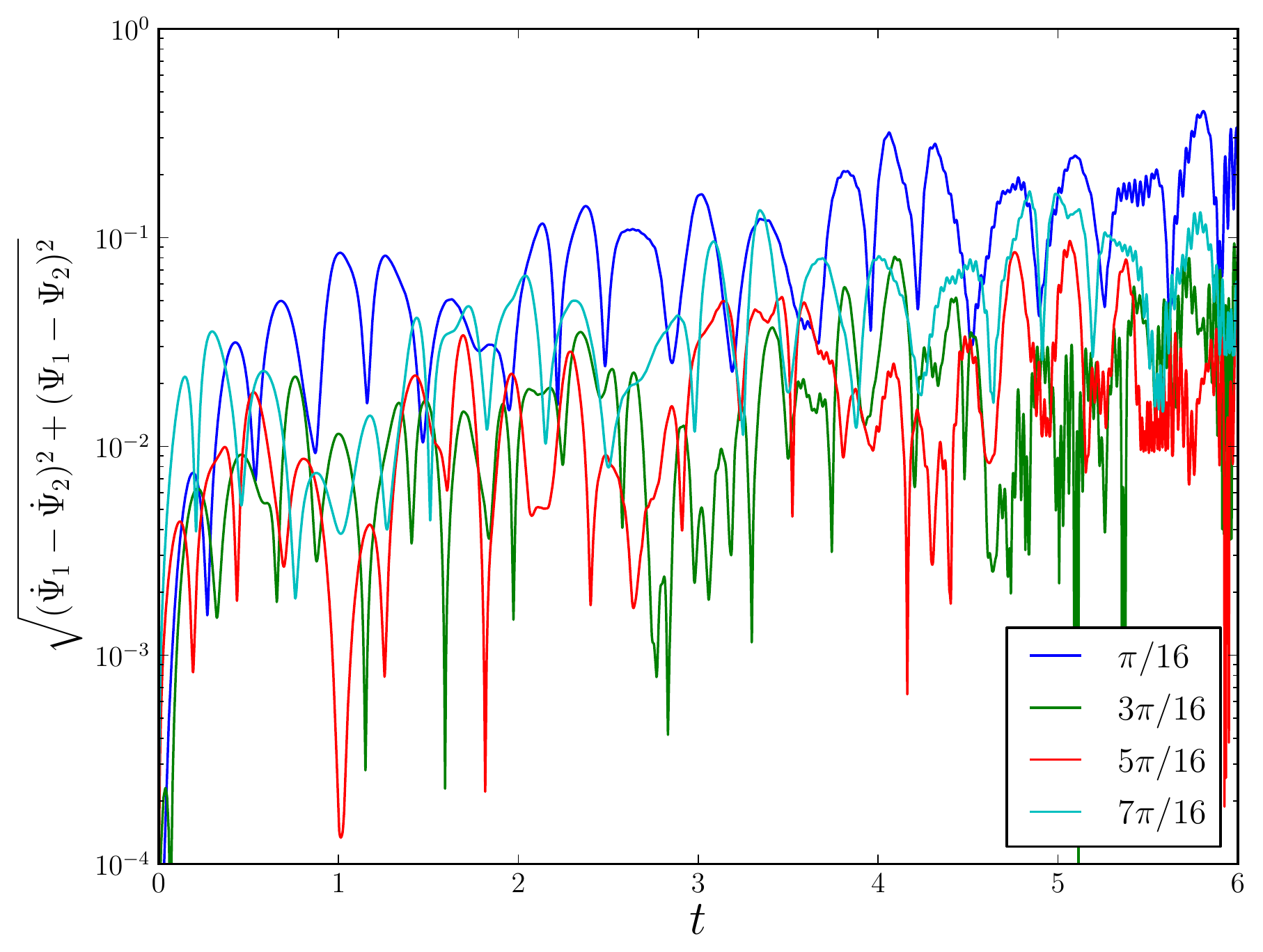}
\caption{\label{LE} For two collapsing simulations with amplitudes $\epsilon=0.2, 0.21$ we plot the phase space distance in logarithmic scale as a function of time. For all points the growth trend implies, according to our definition,  a positive Lyapunov exponent.}
\end{center}
\end{figure}

As further evidence in favor of sensitivity to the initial conditions we consider the largest Lyapunov exponent. We need to face
a number of issues. First the standard definition applies to dynamical system, we are considering a PDE situation. Another problem in applying the standard definition is that we cannot go to asymptotically large times
as we are bounded by the apparent horizon time.  We would
thus naturally fix a point in space and consider the following
quantity {\small $\lambda (T) = \ln \sqrt{(\Psi_{\epsilon_1}(x_0, T)-\Psi_{\epsilon_2}(x_0, T))^2
+ (\dot{\Psi}_{\epsilon_1}(x_0, T)-\dot{\Psi}_{\epsilon_2}(x_0, T))^2}$}, we then define the largest Lyapunov exponent as the
slope of the $(\lambda(T),T)$ graph.  With all the previous caveats mentioned, we have explored the value of this quantity for various points $x_0$'s and have found it to be positive, pointing to exponential sensitivity to the initial conditions. The results are presented in figure \ref{LE}. Note that we have considered two nearby amplitudes $\epsilon=0.2, 0.21$. We will present a more exhaustive discussion of this quantity elsewhere.

{\it  Conclusions } --
Our work can be considered as supporting evidence in favor of the proposal recently put forward  by Hawking \cite{Hawking:2014tga} whereby, even
within the context of the AdS/CFT correspondence it is practically impossible to reliably extract information during gravitational
collapse. We have explicitly presented results indicating that translating information from the gravitationally collapsing bulk to the CFT involves chaotic behavior and
therefore leads to
practical loss of information in the sense that initial uncertainties are exponentially magnified even within a unitary process.

In this manuscript we have considered only part of the phase space of gravitational collapse where an apparent horizon forms in the first few  approaches of the scalar field. It would be interesting to consider a situation where an apparent horizon forms only after a large number of oscillations of the
scalar field; this will potentially make the asymptotic nature of chaos more manifest but will require higher precision. There is evidence that, in general, gravitational collapse has a more complicated phase space than originally assumed in \cite{Bizon:2011gg}. Namely, it has been argued that there are gravitational configurations that are stable against collapse due to the presence of a mass scale \cite{Dias:2012tq}; these claims have been supported by numerical investigations \cite{Buchel:2013uba}. It would be interesting to investigate the effect on these configurations in the context of extracting field theory information.

Let us now discuss the {\it regime of validity} of our calculational framework and its potential extensions.
First, we have used classical gravity in AdS to describe a field theory.
This approximation involves the large $N$ limit in the field theory. Since we avoided regions of strong gravitational curvatures
we refer to field theories with a strong 't Hooft coupling. Second, we have used the KG equation as a description of an operator in the field theory;
this means that we are necessarily referring to operators with relatively small values of the conformal dimension.

To include more general operators would require objects beyond a classical field. For example, operators of very large dimensions are
usually characterized by strings or branes on the gravity side. Chaotic behavior or non-integrability of some classical configurations of strings in the  context
of the AdS/CFT has been recently established for several interesting string theory backgrounds
\cite{Zayasa1,Basu:2011dg,Basu:2012ae,Stepanchuk:2012xi,Basu:2011fw,Basu:2011di,Chervonyi:2013eja,Giataganas:2013dha}. It is very plausible that the
classical string will display chaotic behavior in the background of gravitational collapse.

We can also scrutinize the regime of validity for the KG equation. In the framework of string theory, the KG equation for the scalar field $\Psi$ is
itself an approximation for the corresponding vertex operator. This would correspond to a regime analogous to {\it quantum chaos}; in some
restricted sense
similar problems become tractable (see \cite{PandoZayas:2012ig,Basu:2013uva} for work in the context of confinement). It is expected that in the stringy
treatment of the field $\Psi$ the types of questions that can be formulated change considerably. As opposed to classical chaos, quantum chaos is no longer
 concerned with solutions of the classical equations of motion, their phase space properties and sensitivity to changes in initial
 conditions since such sensitivity does not exist in the quantum case, instead quantum chaos is concerned, for example,  with the statistics of the
 spectrum of energy eigenvalues  \cite{gutzwiller-quantum-chaos}. More importantly, as shown in the seminal study \cite{Chirikov} of the quantum kicked rotor, the quantum behavior can be completely different from the classical one as in dynamical localization. There are also many known cases where
taking the classical limit and the late time behavior do not commute \cite{Ballantine}.

From this point of view it seems that information loss through deterministic chaos is a property of the classical limit of the AdS/CFT correspondence but not of the full quantum correspondence. The restoration of unitary  does not go through as small corrections to a classical picture
but as a drastic reformulation of the problem just as in the case of classical versus quantum chaos.

{\it Acknowledgments.} --  We thank R. Akhoury, L. Bieri, D. Garfinkle, A. Hashimoto, C. Keeler, J. Liu , H. P. de Oliveira and E. Rasia for comments and suggestions. We are particularly thankful to D. Reichmann for insightful criticism and suggestions. All simulations were performed in the University of Michigan Flux high-performance computing cluster. This work is  partially supported by
Department of Energy under grant DE-FG02-95ER40899.

\end{document}